%% file: bruzual_paper.tex
\documentstyle[rrla,psfig]{article}
\begin{document}
\title{A REVISED AGE FOR THE z=1.55 GALAXY LBDS 53W091}

\author{Gustavo Bruzual A. and Gladis Magris C.\altaffilmark{1}\\
bruzual@cida.ve, magris@cida.ve}
\altaffiltext{1}{Centro de Investigaciones de Astronom{\'{\i}}a (CIDA),
Apartado Postal 264, M\'erida, Venezuela}

\begin{abstract}
Empirical evidence suggests that the stellar population in LBDS 53W091 is
significantly younger than that in M32. Population synthesis models indicate
that the age of LBDS 53W091 is in the range from 1 to 2 Gyr and depends on the
specific model. Older ages require sub-solar metallicity models.
The estimates of the age of the dominant population in M32 range from 3 to
5 Gyr and depend not only on
the model but also on the SED of this galaxy used in the fits.
The same models predict an age of 11 to 13 Gyr for the stars in a typical old
E/S0 galaxy.
A 1 to 2 Gyr old galaxy at $z = 1.55$ poses no problem for the
$\Lambda = 0,~\Omega = 1$ universe, as long as $h \le 0.8$.
The most likely reason for the difference between our age estimate
and the value of 3.5 Gyr derived by Dunlop et al. (1996), is 
the fact that these authors did not require that the population synthesis models
that they used fitted simultaneously the UV break amplitudes and the observed
$R-J,~R-H$, and $R-K$ colors of this galaxy.
As we will discuss elsewhere, a comparison of our results with those by Spinrad
et al. (1997) shows that our conclusions still hold.
The paper by Spinrad et al. was unknown to us at the time of this Conference.
\end{abstract}

\section{Introduction}

LBDS 53W091, at $z = 1.552$, is one of the reddest and most distant galaxies
with high quality UV spectrophotometry and UV-optical photometry.
The discovery spectrum by Dunlop et al. (1996) samples the near UV in the
galaxy rest frame and shows clearly several absorption features and the
spectral breaks at $\lambda_{rest} = 2640$ and 2900 \AA. Observer frame
$RJHK$ photometry was also performed by Dunlop et al.
{From} a comparison of the amplitude of these breaks with the predictions of various
population synthesis models (Guiderdoni \& Rocca-Volmerange 1987; Bruzual \& Charlot 1993;
Bressan, Chiosi, \& Fagotto 1994; and their own models),
Dunlop et al. estimate that star
formation ceased in this galaxy at least 3.5 Gyr before the epoch of the
observation and use this number to constrain the value of the density parameter
$\Omega$. For the stellar population in this galaxy not to be older than the
($\Lambda = 0$) universe at this epoch,
$\Omega < 0.2$ if $h \equiv H_0/100$ km s$^{-1}$ Mpc$^{-1} = 0.75$, or
$\Omega < 0.8$ if $h = 0.50$. Since the age of the $\Omega = 1$ universe at
$z=1.552$ (1.6$h^{-1}$ Gyr) is $< 3.5$ Gyr for the accepted range of values of
$h$ ($0.50 < h < 0.75$), Dunlop et al. argue that the $\Lambda = 0,~\Omega = 1$
universe can thus be {\it formally excluded} as a valid world model.

\input fig1

Given the far reaching astrophysical implications of this conclusion, we
repeat Dunlop et al. (1996) analysis of their data in an independent
manner. We show below that the age of the stellar population in LBDS 53W091 is
most likely in the range 1 to 2 Gyr.
Thus, the $\Lambda = 0,~\Omega = 1$ universe is {\it still viable}, as long as
$h \le 0.8$.

\section{Observational Data}

In Table 1 we summarize the photometric data available for LBDS 53W091.
The $RJHK$ magnitudes were taken from Dunlop et al. (1996).
The $V$ and $I$ magnitudes were derived from the synthetic $V-R$ and
$R-I$ colors computed from the Dunlop et al. spectral energy distribution (SED),
kindly made available to us by H. Spinrad, D. Stern, and A. Dey.
In Table 1, $\lambda_{eff}$ and $\Delta\lambda$ are the effective wavelength
and the range covered by each filter when observing a source at $z=1.552$.
$\beta(2640)$ and $\beta(2900)$, defined in \S4.2, are not expressed in mag.

As Dunlop et al. did, we also compare the SED of LBDS 53W091 with 
that of the dwarf elliptical galaxy M32. Additionally, we compare both of these
spectra with the SED of a giant E/S0 galaxy, typical of an old stellar
population. In Fig 1 the Dunlop et al. (1996) spectrum in the UV below 3700 \AA\ 
is shown as a histogram.
The values of the flux in the $R, J, H$, and $K$ bands were computed from the
data in Table 1 and are shown as individual points at $\lambda_{eff}$.
The size of the error bars correspond to $\pm \sigma$ in Table 1.
Including the additional uncertainty introduced by matching the large aperture
photometry to the SED (at $R$) will increase the size of the error bars.
The SEDs of M32 and the E/S0 galaxy given by Bica et al.
(1996; E7-M32 and E2-E5 groups in their notation) are also shown in Fig 1.
The SED of M32 used by Magris \& Bruzual (1993), marked M32(b), is redder than
the SED from Bica et al. This difference is due to the uncertainty in matching
UV and optical SEDs. These empirical data already
suggest that LBDS 53W091 is considerably younger than M32.

\section {Population Synthesis Models}

Bruzual \& Charlot (1997, hereafter BC97; see also Bruzual \& Charlot 1996, and
Bruzual et al. 1997) have extended their (1993)
evolutionary population synthesis models to provide the evolution in time of
the spectrophotometric properties of simple stellar populations (SSPs) of a
wide range of stellar metallicity ($Z = 0.0001, 0.0004, 0.004, 0.008, 0.02,
0.05, 0.10$). The evolutionary tracks are based on the work from the Padova
School and an independent set of tracks from the Geneva School for
$Z_\odot = 0.02$. The library of synthetic stellar SEDs compiled by Lejeune
et al. (1997a,b, hereafter LCB) that covers all the metallicities listed above
is used by BC97.
The nearly-empirical, extended version of the Gunn \& Stryker (1983,
hereafter EGS) stellar atlas of BC93 has been updated and used to build
the reference $Z_\odot$ models. In the EGS atlas the IUE stellar library
of Fanelli et al. (1996) has been incorporated, and the spectra of the coolest
M-type stars have been replaced following LCB.
Thus, for $Z = Z_\odot$ we can compare the predictions of purely-theoretical
models (built with LCB) and nearly purely-empirical models (built with
EGS) for either the Padova or the Geneva tracks.

\input fig2

\input table1

\input table2

\section{The Age of LBDS 53W091}

\subsection{Age from Broad Band Colors}

Fig 2 shows the evolution in time of various colors computed in the observer
frame, assuming $z=1.552$, for a SSP and a 1 Gyr burst model (1GB hereafter).
In the latter the star formation rate (SFR) is constant during the first Gyr
and zero afterward. The colors computed from the SEDs shown in Fig 1 have been placed at
the age at which the model produces the best fit to the observed colors
(Table 2 and Fig 3).
The height of the boxes representing the colors of LBDS 53W091
correspond to $2\sigma$ (Table 1). The width of these boxes represents
the range of age allowed by the model in each color.
The run of $m(\lambda) - R$ vs $\lambda$ for the three
galaxies and the SSP model is shown in Fig 3. The fit of the model to the
observed colors at the indicated age is excellent for the three galaxies.
Similar fits are obtained for different models at slightly different ages
(Table 2).
Models built with the Geneva tracks predict younger ages for the E/S0
galaxy than the Padova track models. The Padova tracks include overshooting in
the convective core of stars more massive than $1\ M_{\odot}$ whereas the
Geneva tracks stop overshooting at $1.5\ M_{\odot}$. Stars in this mass
range require more time to leave the MS in the Padova tracks than in the Geneva
tracks. Hence, the Geneva track models become red faster.

{From} this analysis we conclude that:
{\it (a)}
There is a significant difference in the age of the stellar population
dominant in M32 and LBDS 53W091 amounting to over 1 Gyr if $Z = Z_\odot$.
For the same SFR and metallicity, the ages obtained from either Salpeter (1955)
or Scalo (1986) IMF models are not significantly different.
The age difference increases to over 2 Gyr if the M32(b) SED is used.
{\it (b)}
The observed colors of LBDS 53W091 are consistent with those predicted
by $Z = Z_\odot$ models at an age from 1 to 2 Gyr, depending on the SFR,
the IMF, and the evolutionary tracks.
{\it (c)}
The age derived for the M32(b) SED using the BC97 models is consistent with
that obtained by Magris \& Bruzual (1993) using the BC93 models.
This result establishes the consistency of both sets of models.
{\it (d)}
Considerably older ages are predicted for LBDS 53W091 and M32 if we assume
that the metallicity of the dominant population in these galaxies is
$Z = Z_\odot/5$ (Table 2).
These models never become as red as the M32(b) and E/S0 SEDs.
{\it (e)}
The age determined from the 1GB models is less than 1 Gyr older
than the age determined from the SSP models. As expected, this age difference
decreases as the age of the galaxy increases and the youngest stars in
the 1GB model approach the lower main sequence (MS).

\input fig3

\subsection{Age from the UV Spectrum}

The amplitude of the spectral breaks at $\lambda = 2640$ and 2900 \AA\ is
a function of the age of the stellar population.
However, the actual value of the amplitude of the break and of its rate of
change depend on the wavelength intervals used to measure the flux above and
below the discontinuity.
Experimenting with several wavelength intervals we found that the 30 \AA\ bins
used by Dunlop et al. (1996) to define these amplitudes are too narrow (just a single 
wavelength point is included in some instances),
resulting in break amplitudes with
little sensitivity to the evolution of the stellar population.
Instead, we define the following spectral properties which provide more sensitivity
to the evolution of the population than the break amplitudes defined by Dunlop et al.
$$ 
{
\beta(2640)= {
{(\lambda_4-\lambda_3)^{-1}\int_{\lambda_3}^{\lambda_4}F_\lambda(\lambda)d\lambda}
\over
{(\lambda_2-\lambda_1)^{-1}\int_{\lambda_1}^{\lambda_2}F_\lambda(\lambda)d\lambda}
}
}
,\ \ \ \ \ \ \ 
{
\beta(2900)= {
{(\lambda_6-\lambda_5)^{-1}\int_{\lambda_5}^{\lambda_6}F_\lambda(\lambda)d\lambda}
\over {F_\lambda(2800)}
}
},
$$ 
where
$(\lambda_1, \lambda_2)$ = (2200, 2400),
$(\lambda_3, \lambda_4)$ = (2640, 2750), and
$(\lambda_5, \lambda_6)$ = (2900, 3150) \AA.
$F_\lambda(2800) = F_\lambda$(Mg {\sc II} 2800)
measures the flux at the bottom of the Mg {\sc II} absorption line at 2800 \AA.

\input fig4

Fig 4 shows the evolution of $\beta(2640)$ and $\beta(2900)$ for the same models
used in Fig 2. The dotted lines join the points corresponding to individual
stars at the MS turnoff.
The dots representing M32 and the E/S0 galaxy have been placed at
the age at which the model produces the best fit to the observed colors
(Fig 3). The height of the boxes representing LBDS 53W091 correspond to
$2\sigma$. The width of these boxes represents the range of age allowed by
the model. The triangles, from left to right, represent the
position of stars of 1.8, 1.5, 1.3, 1.15, and 1 $M_\odot$.
The effective temperature at the turnoff was obtained from the Padova tracks.
{From} this figure we conclude that:
{\it (a)}
$\beta(2900)$ is a good age indicator for young stellar populations
up to 3 Gyr. LBDS 53W091 and M32 fall on top of the model prediction at the
age determined from fitting the colors.
For LBDS 53W091, $\beta(2900)$ indicate again an age
in the range from 1 to 2 Gyr, considerably younger than M32.
{\it (b)}
The predicted $\beta(2640)$ agrees very well with the observed
values for M32 and E/S0 at the age derived from broad band colors.
Because of its flatter slope, $\beta(2640)$ is not as
useful an age indicator as $\beta(2900)$. Despite the wide range of possible
age, the value of this break for LBDS 53W091 indicates that this galaxy is
younger than M32.
{\it (c)}
The values of $\beta$ for the integrated populations are higher at early
ages than the corresponding value for the star at the
MS turnoff (dotted lines in Fig 4).
The contribution from stars cooler than the turn-off
star is noticeable even in the near UV. This happens because $\beta$
increases with decreasing stellar mass and low mass stars are present in large
numbers in the MS according to any realistic IMF.
The age derived from $\beta(2900)$ for LBDS 53W091 is about 1.5 Gyr younger
than the MS turnoff for the same amplitude of this break.
At the age of M32, the age of the population agrees fairly well with the
turnoff. {\it (d)} The E/S0 galaxy shows a value of $\beta(2900)$ higher than
predicted by the models.
All the dwarf and giant stars of type later than G8 in the Fanelli
et al. (1996) IUE library (used by BC97) show chromospheric emission at
Mg {\sc II} 2800, filling in the the absorption line, and decreasing $\beta(2900)$
with time in the models. In real E/S0 galaxies chromospheric emission must also occur,
but most likely there is a distribution of emission line intensity at each
spectral type, not taken into account in our models. This fact may explain the
apparent discrepant value of $\beta(2900)$ for the E/S0 galaxy in Fig 4.

\subsection{Age from the Galaxy SED}

Fig 3 of Bruzual \& Magris
(1997; hereafter Fig 3bm), not included here due to lack of space,
shows the SEDs for the $Z = Z_\odot$ SSP model together with the
observed SEDs. The ages derived from fitting broad band colors agree with the
ages derived from fitting the complete SEDs from 2000 to 9600 \AA.
The quality of the fits in Fig 3bm is remarkable.
We also show in Fig 3bm the contribution from different stellar groups
to the total SED. We see from this figure
that despite the relatively similar value of $\beta(2640)$ for the 3 galaxies
(Fig 4), the relative contribution of the MS and subgiant branch stars is quite
different for each galaxy. It is important to use models that include all
phases of stellar evolution to study these systems.

\section{Conclusions}

Empirical evidence suggests that the stellar population in LBDS 53W091 is
significantly younger than the dominant population in M32 (Fig 1).
We have used evolutionary population synthesis models to estimate the age
of the dominant population in these stellar systems.
The age of LBDS 53W091 is in the range from 1 to 2 Gyr and depends on the
specific model (Table 2). Older ages require sub-solar metallicity models.
The estimates of the age of the dominant population in M32 range from 3 to 5 Gyr
and depend not only on
the model but also on the SED of this galaxy used in the fits.
The same models predict an age of 11 to 13 Gyr for the stars in a typical old
E/S0 galaxy. This age is consistent with the age of the metal-rich
galactic bulge globular clusters NGC 6553, NGC 6528, and Terzan 5, estimated
by Bruzual et al. (1997).

Passive evolution seems an adequate scenario for the evolution of the stellar
population in E galaxies from $z = 1.6$ to $z=0$.
The dominant population in M32 is genuinely young (3 to 5 Gyr), independently
than an older stellar population may be present in this galaxy.
Thus, M32 may not be representative of galaxies that evolve passively.
The length of time for which these galaxies have existed as individual
dynamical entities is not determined by our models.

The age of the $\Lambda = 0,~\Omega = 1$ universe at $z=1.552$ is 1.6$h^{-1}$
Gyr. Hence, a 1 to 2 Gyr old galaxy at $z = 1.552$ poses no problem for this
universe as long as $h \le 0.8$. In this universe, the 13 Gyr limit for the E/S0
galaxy at $z=0$ requires $h \le 0.5$.

The most likely reason for the difference between our age estimate of 1 to 2 Gyr for
LBDS 53W091 and the value of 3.5 Gyr derived by Dunlop et al. (1996), is
the fact that these authors did not require that the population synthesis models
that they used fitted simultaneously the UV break amplitudes and the observed
$R-J,~R-H$, and $R-K$ colors of this galaxy.
When we prepared the poster for this Conference the paper by Spinrad et al.
(1997) was not known to us. A preliminary comparison of our results with those
by Spinrad et al. shows that our conclusions still hold. 
This will be discussed in detail elsewhere.

We thank H. Spinrad, A. Dey, and D. Stern for kindly sending to us
their remarkable spectrum of LBDS 53W091.

\input appendix

\end{document}

%% file: fig1.tex
\begin{figure}
\vspace*{70mm}
\begin{minipage}{70mm}
\includegraphics{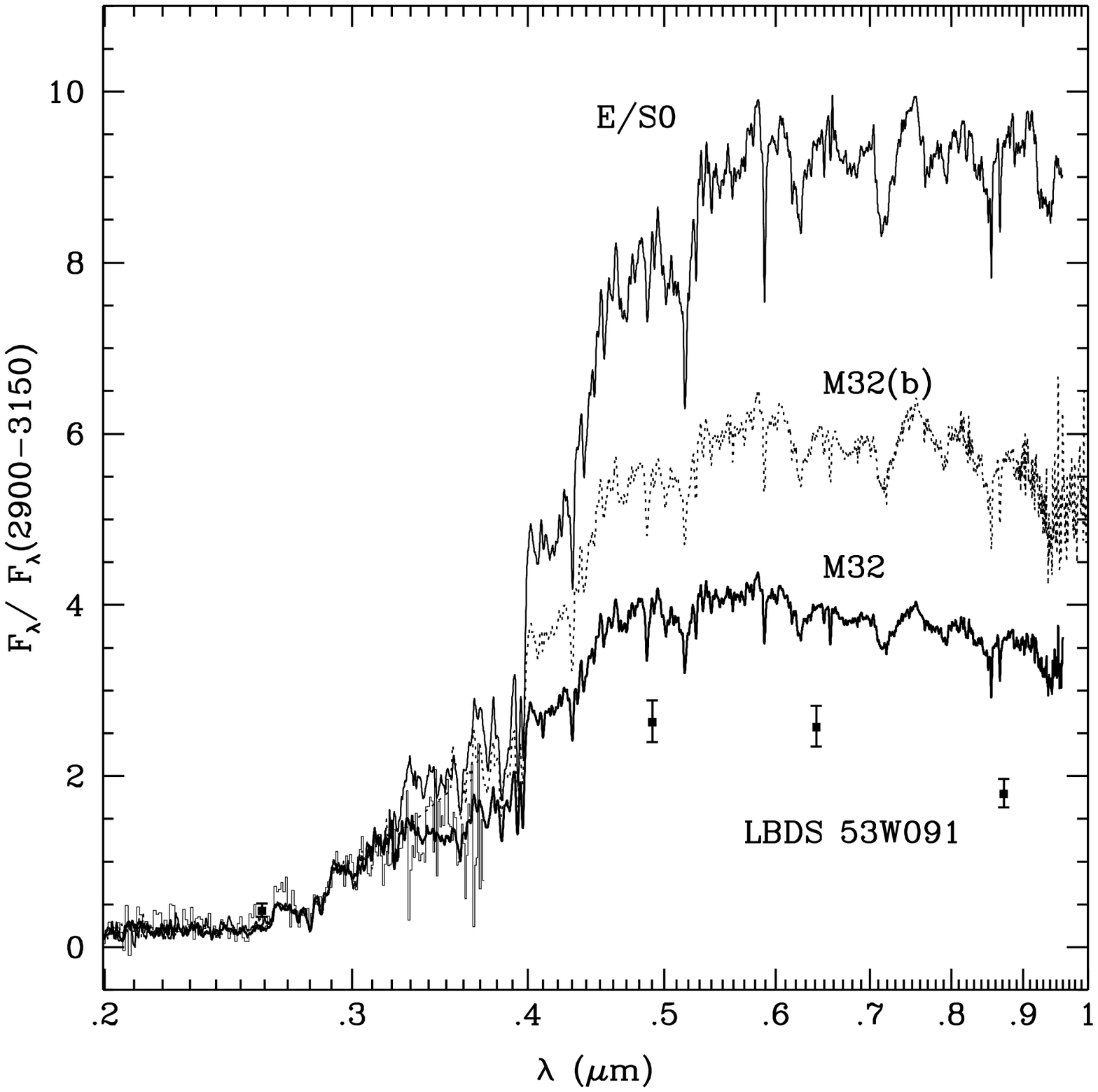} 
\end{minipage}
\caption{SEDs of LBDS 53W091, M32 and a giant E/S0 in the galaxy rest frame.
The SEDs are scaled to the same value of the integrated flux in the range from
2900 to 3150 \AA. The wavelength scale is logarithmic.}
\end{figure}

%% file: fig2.tex
\begin{figure}
\vspace*{90mm}
\begin{minipage}{90mm}
\includegraphics{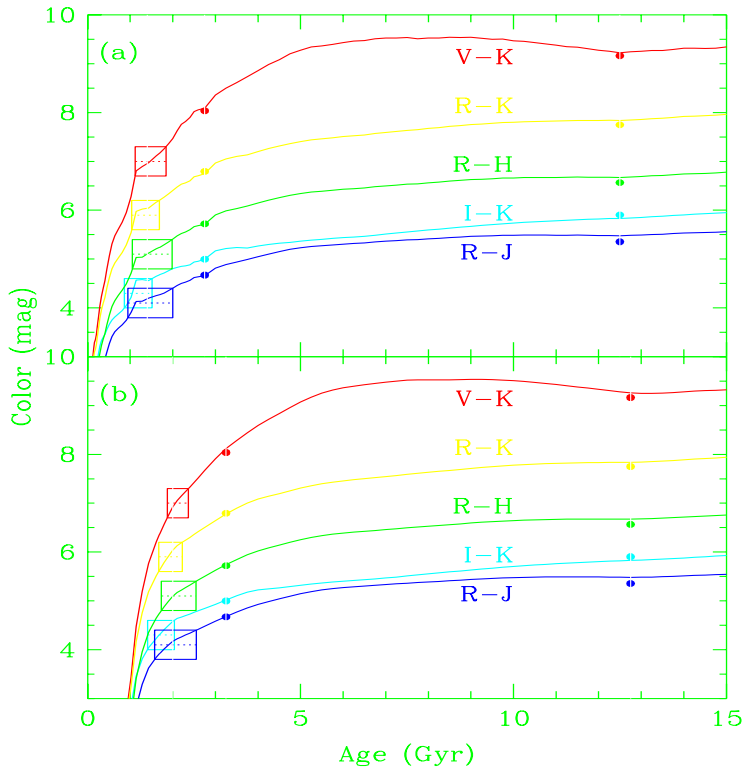} 
\end{minipage}
\caption{
Evolution in time of various colors computed in the observer frame assuming
$z=1.552$ for: {\it (a)} a SSP model, and {\it (b)} a 1GB model.
Both models were computed with the $Z = Z_\odot$ Padova tracks, using
the EGS atlas and the Salpeter (1955) initial mass function (IMF).
}
\end{figure}

%% file: table1.tex
\begin{table}[h]
\caption{Photometric Data }
\label{tab1}
\begin{center}\leavevmode
\begin{tabular}{cccccccc}\hline
Band & mag & $\lambda_{eff}$ & $\Delta\lambda$ & Band & mag & $\lambda_{eff}$ & $\Delta\lambda$ \\ \hline
$V$ & 25.7 $\pm$ 0.2 & 2157 & 1861-2900 & $H$ & 19.5 $\pm$ 0.1 & 6413 & 5251-7284 \nl
$R$ & 24.6 $\pm$ 0.2 & 2588 & 2175-3429 & $K$ & 18.7 $\pm$ 0.1 & 8716 & 7445-9948 \nl
$I$ & 23.0 $\pm$ 0.2 & 3096 & 2743-3487 & $\beta(2640)$  & 1.9 $\pm$0.3 & \nodata & $(2640:2750) \over (2200:2400)$ \nl
$J$ & 20.5 $\pm$ 0.1 & 4904 & 4075-5643 & $\beta(2900)$  & 4.1 $\pm$0.4 & \nodata & $(2900:3150) \over 2800$        \nl \hline
\end{tabular}
\end{center}
\end{table}

%% file: table2.tex
\begin{table}[h]
\caption{Galaxy Age Derived from Fits to Different Models (Gyr)}
\label{tab2}
\begin{center}\leavevmode
\begin{tabular}{cclcccccc} \hline
Tracks & SFR & IMF & $Z$ & Atlas & 53W091 & M32 & M32(b) & E/S0 \\ \hline
Padova & SSP & Salpeter & $Z_\odot$ & EGS &  1.4 & 2.75 & 3.50 & 12.50 \nl
   "   &  "  & Scalo    &     "     &  "  &  1.4 & 3.00 & 3.50 & 12.50 \nl
   "   & 1GB & Salpeter &     "     &  "  &  2.0 & 3.25 & 4.00 & 12.75 \nl
   "   &  "  & Scalo    &     "     &  "  &  2.1 & 3.50 & 4.25 & 13.00 \nl
Geneva & SSP & Salpeter &     "     &  "  &  1.2 & 3.25 & 4.50 & 11.25 \nl
   "   &  "  & Scalo    &     "     &  "  &  1.3 & 3.50 & 4.75 & 11.00 \nl
   "   & 1GB & Salpeter &     "     &  "  &  2.0 & 3.75 & 5.00 & 11.50 \nl
   "   &  "  & Scalo    &     "     &  "  &  2.1 & 4.00 & 5.25 & 11.75 \nl
Padova & SSP & Salpeter &$Z_\odot/5$& LCB & 3.75 & 13.50 & \nodata & \nodata \nl
   "   &  "  & Scalo    &     "     &  "  & 4.00 & 13.75 & \nodata & \nodata \nl
   "   & 1GB & Salpeter &     "     &  "  & 4.50 & 14.00 & \nodata & \nodata \nl
   "   &  "  & Scalo    &     "     &  "  & 4.50 & 14.00 & \nodata & \nodata \nl \hline
\end{tabular}
\end{center}
\end{table}

%% file: fig3.tex
\begin{figure}
\vspace*{90mm}
\begin{minipage}{90mm}
\includegraphics{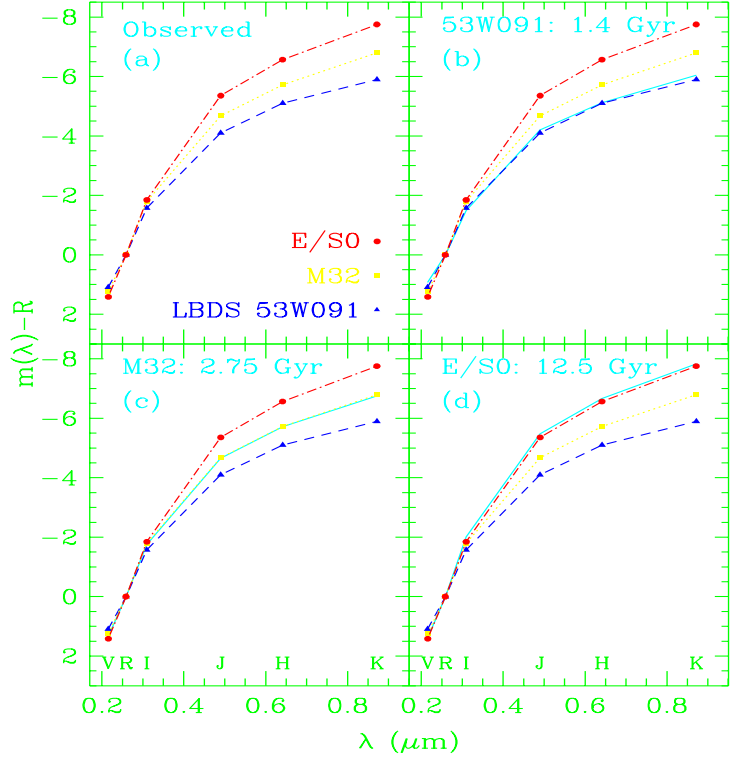} 
\end{minipage}
\caption{
{\it (a)} $m(\lambda) - R$ vs $\lambda$ for LBDS 53W091 (dashed line),
M32 (dotted line), and E/S0 (dot-dashed line).
{\it (b,c,d)} Colors predicted by the SSP model (Salpeter IMF, EGS atlas), at
1.4, 2.75, and 12.5 Gyr, respectively (solid line).
The effective wavelengths seen by the $VRIJHK$ filters for $z = 1.552$
are indicated in the two bottom frames.
}
\end{figure}

%% file: fig4.tex
\begin{figure}
\vspace*{90mm}
\begin{minipage}{90mm}
\includegraphics{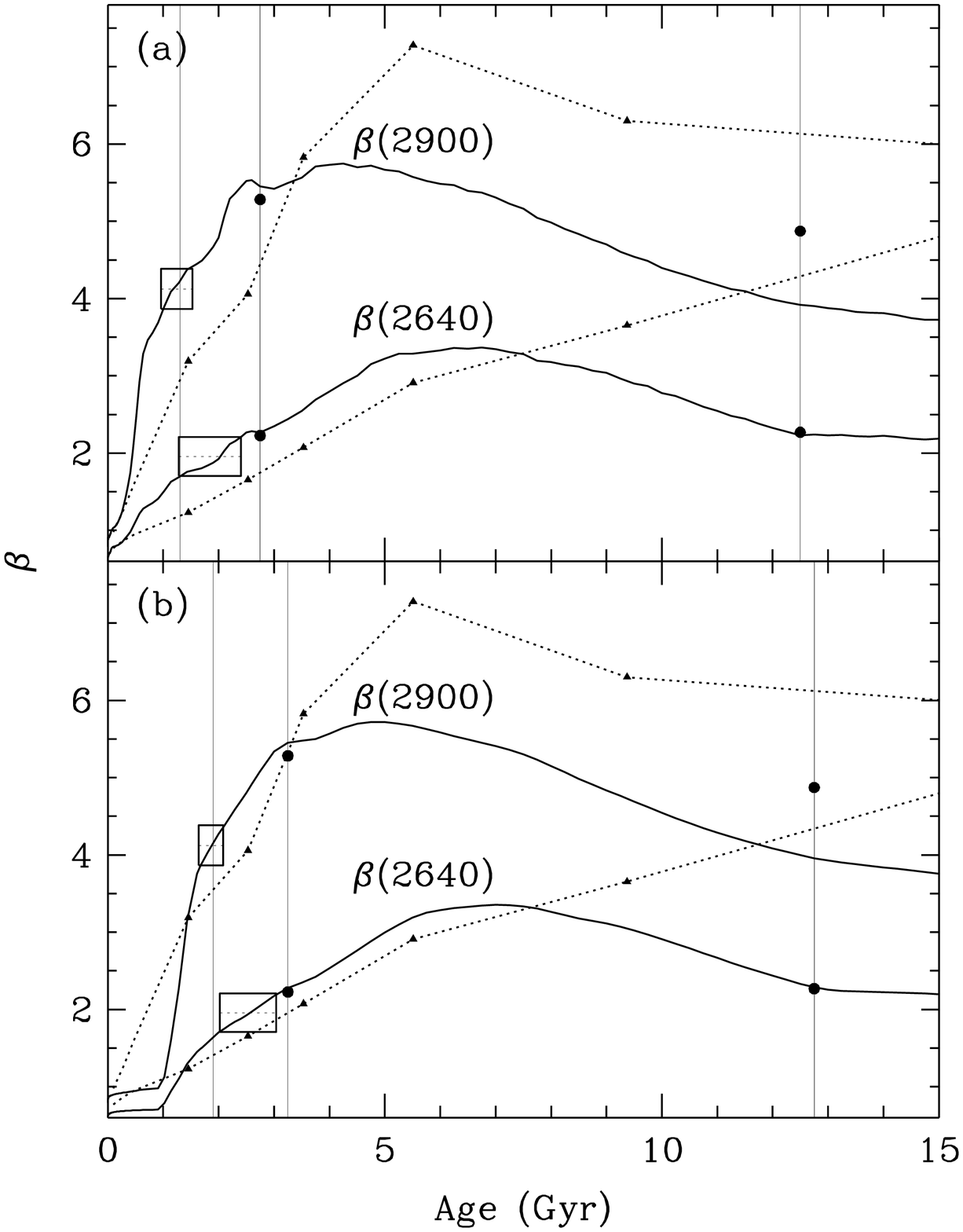} 
\end{minipage}
\caption{
Evolution in time of $\beta(2640)$ and $\beta(2900)$ for: {\it (a)} a SSP model,
and {\it (b)} a 1 Gyr Burst model, both for the $Z = Z_\odot$ Padova tracks
and the Salpeter IMF.
}
\end{figure}

%% file: appendix.tex
\input fig5
\section{Appendix}

For the benefit of the interested reader,
in Fig 5 of this Appendix we reproduce Fig 3bm, i.e., Fig 3 from Bruzual \& Magris
(1997) together with the corresponding discussion.

Fig 5 shows the SEDs for the $Z = Z_\odot$ SSP model together with the
observed SEDs. The ages derived from fitting broad band colors agree with the
ages derived from fitting the complete SEDs from 2000 to 9600 \AA.
The quality of the fits in Fig 5 is remarkable.
The observed SEDs have been scaled to match the model
flux by minimizing the residuals $log~F_\lambda(obs) - log~F_\lambda(model)$
over the entire SED. The resulting fluxes correspond to a galaxy of mass
1 $M_\odot$.
To illustrate the effects of spectral evolution more clearly, the SEDs in panels
(b) and (c) have been multiplied by a scaling factor $\alpha$.
The value of $\alpha^{-1}$
reflects the amount of spectral evolution from 1.4 to 12.5 Gyr. Thus, at 5500
\AA, the M32 model is half as bright as the LBDS 53W091 model, and the
E/S0 model is 7 times fainter than LBDS 53W091.
These numbers refer only to the dimming effects of passive evolution
in galaxies of identical mass. No attempt has been made to scale the flux
according to the mass of each galaxy.
In each panel of Fig 5 the contribution of each stellar group to the total
SED is shown as follows:
main sequence (dotted line),
subgiant branch (short-dashed line),
red giant branch (long-dashed line),
horizontal branch (dot and short-dashed line),
asymptotic giant branch (dot and long-dashed line).
Despite the relatively similar value of $\beta(2640)$ for the 3
galaxies (Fig 4), the relative contribution of the MS and subgiant branch
stars differs for each galaxy.
It is important to use models that include all
phases of stellar evolution to study these stellar systems.

%% file: fig5.tex
\begin{figure}
\vspace*{90mm}
\begin{minipage}{90mm}
\includegraphics{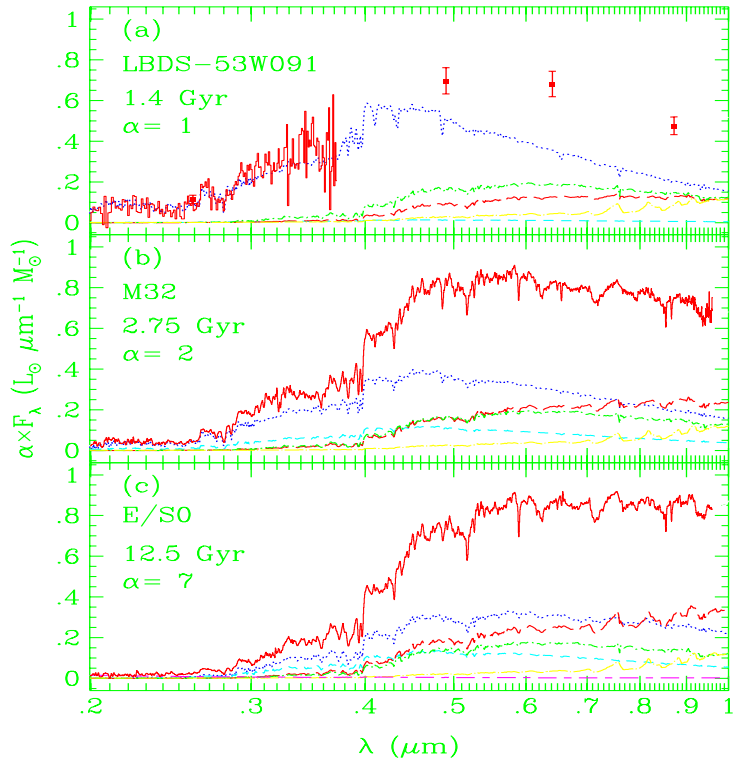} 
\end{minipage}
\caption{
{\it (a,b,c)} SED for the $Z=Z_\odot$ SSP model at 1.4, 2.75, and 12.5 Gyr (heavy
solid lines and points ) together with the SED of LBDS 53W091 in the galaxy rest
frame, M32, and the E/S0 galaxy (light solid lines).
}
\end{figure}